\documentstyle[11pt,moriond,epsfig]{article}
\input amssym.def
\input amssym.tex

\def\be{\begin{equation}}
\def\ee{\end{equation}}
\def\bea{\begin{eqnarray}}
\def\eea{\end{eqnarray}}

\begin{document}
\title{Dipoles at $\nu =1$}
\author{V. PASQUIER}
\address{CEA/Saclay, Orme des Merisiers,\\
B\^{a}t. 774, F-91191 Gif-sur-Yvette Cedex, FRANCE}
\maketitle
\abstracts{
We consider the problem of Bosonic particles interacting repulsively in a
strong magnetic field at the filling factor $\nu =1.$\ We project the system
in the Lowest Landau Level and map the dynamics into an interacting Fermion
system.\ We study the resulting Hamiltonian in the Hartree--Fock
approximation in the case of a $\delta $ repulsive potential.\ The physical
picture which emerges is in agreement with the proposal of N. Read that the
composite Fermions behave as a gas of dipoles.\ We argue that the
consequence of this is that the composite Fermions interact with screened
short range interactions.\ We develop a Landau theory which we also expect
to describe the physical $\nu =1/2$ Fermionic state. The Form factor, the
effective mass and the conductivity are analised in this model.}

\section{Introduction\protect\footnote{%
This lecture is issued from an unpublished paper on $\nu =1$ Bosons
(``Composite fermions and confinement'' March 1996).
Although some results are outdated,
some aspects considered here have not been discussed
in the recent literature.
I give a list of recent references on the subject for the
interested reader [11-15].}}

There has been a renewed interest in the quantum Hall effect when
the filling factor is a fraction with an even denominator.\ Willets and his
collaborators \cite{willet} have observed an anomalous behavior in the
surface acoustic wave propagations near $\nu =1/2$ and $\nu =1/4$.\ A
remarkable outcome of their experiments is that they probe a longitudinal
conductivity $\sigma _{xx}\left( q,\omega \right) $ increasing linearly with
the wave vector $q$.\ Halperin, Lee and Read \cite{halperin} have suggested
that the system exhibits a Fermi liquid behavior at this particular value.\
Another approach followed by Rezayi and Read \cite
{rezayi} and Haldane et al.\ \cite{haldane} consists in obtaining trial wave
functions which enable to study numerically the properties of the system at
this filling factor.\ In these studies the cyclotron frequency is supposed
to be sufficiently large so that the only relevant excitations are confined
to the lowest Landau level.
Here we introduce a model which accounts for the success
of these trial wave functions and enables to compute the physical properties
of the system in the infinite cyclotron energy limit.
This is a different approach than the Chern Simon field theory
which mixes the Landau levels and requires the mass of the electron as a
parameter of the theory.\ We do not address the physical problem of
electrons in a magnetic field at $\nu =1/2$.\ Instead we have considered the
problem of Bosonic particles interacting repulsively in a magnetic field at
a filling factor $\nu =1$.\ Although it may at first look quite different,
the problem of formation of a Fermi sea is
essentially the same as in the $\nu =1/2$ case.\ If one applies the analyses
of the composite Fermions or the Chern Simon approach to such a system, one
is essentially led to the same picture of Fermi sea formation as in the $\nu
=1/2$ case.\ We have also verified this hypotheses by performing a numerical
simulation for a small system on a sphere (see figure 1).\ The main reason
why this is simpler theoretical problem to look at than the $\nu =1/2$
physical problem is that the wave function one needs to start from is the
Slater determinant of the lowest Landau level one body wave functions which
is a much simpler object to consider than the Laughlin $\nu =2$ wave
function which one would have to use in the $\nu =1/2$ case.

Read has interpreted the fluxes attached to the electron as physical
vortices bound to it \cite{read}.\ We believe that his proposal differs
considerably from the mean field interpretation for the following reason.\
The mean field treats the composite electron as a charged particle which
couples minimally to the electro-magnetic field.\ In Reads picture, the
vortices carry a charge equal to minus one half of that of the electron so
that the bound state must be viewed as a neutral particle which propagates
in a constant charge background.\ In this case the response to an electric
field depends on the internal structure of the composite object.
We are led to this picture in the $\nu =1$ case.\ The
essential simplification is that there is a single vortex coupled to the
Boson, this vortex is a Fermion carrying the opposite charge as the Boson
and we can use a second quantized formalism to analyse the model.\ The bound
state is then a dipole whose structure was discovered long ago \cite
{gor'kov}, \cite{lerner}, \cite{kallin}.

Away from $\nu=1/2$, the vortices no longuer carry the opposite charge as the
Fermion and we expect the dipoles to carry an electric charge proportional
$\Delta B$ the difference of the magnetic field with $B_{1/2}$.

\section{The Microscopic Model}

We consider $N$ particles of identical charge interacting with a repulsive
force in a domain of area $\Omega $ thread by a magnetic field $B$
so that the flux per unit area is equal to one.\ We take units where $%
\hslash =1$ and the magnetic length $l=\sqrt{\hslash c/eB}=1$. 
We assume that 
the dynamic can be restricted to
the Lowest Landau Level.\ The one body Hamiltonian has $N$ degenerate
eigenstates, thus in the case where the particles are Fermions the only
accessible state is given by the Slater determinant of the one body wave
functions.\ This state will define the vacuum of the theory.\ We now discuss
the case where there are two sets of particles obeying distinct statistics.\
The first set contains $N_{1}$ Fermions and the second set contains $N_{2}$
Bosons.\ We keep the sum $N_{1}+N_{2}=N$ fixed so that the filling factor
remains equal to one.\ We also keep the interaction equal between all the
particles.\ It is instructive to first look at the case of 1 Boson
interacting with $N-1$ Fermions.\ By performing a particle hole
transformation on the Fermions, we can equivalently regard this as a Boson
interacting with a hole.\ This problem has been studied by Kallin and
Halperin \cite{kallin}.\ A surprising outcome is that the wave functions
which describe this two body state are independent of the potential and are
given by the ground state eigen-functions of the free Hamiltonian \cite
{gor'kov}, \cite{lerner}, \cite{kallin}. They describe a neutral 
dipole with a momentum perpendicular
to its canonical momentum.
Our hypothesis is that the system where several Bosons interact with the same
number of holes reorganizes into neutral fermionic dipoles with a
small residual interaction.
In order to test this hypotheses we have
also performed a numerical evaluation of the ground state energies for a
mixed system consisting of $N_{1}$ Bosons and $N_{2}$ Fermions on a sphere
keeping $N_{1}+N_{2}=N$ fixed.\ The sphere has $N-1$ quantum fluxes and the
interaction between the particles is a delta function interaction.\ The
gross features of the spectrum are those of a system of $N_{2}$ free
Fermions on a sphere with no magnetic field (see fig. 2).

In order to study the system 
we introduce a second quantization formalism by defining
Bosonic $\left( a_{s}^{+}\right) $ and Fermionic $\left( b_{s}^{+}\right) $
creation operators which create the one body states in the Lowest Landau
level.\ They obey the standard commutation relations $\left[
a_{s},a_{s^{\prime }}^{+}\right] =\delta _{s,s^{\prime }},\left\{
b_{s},b_{s^{\prime }}^{+}\right\} =\delta _{s,s^{\prime }}$.\ The vacuum $%
\left| 0\right\rangle $ is the filled Landau level state which is
characterized by $a_{s}\left| 0\right\rangle =b_{s}^{+}\left| 0\right\rangle
=0$.\ We also define the fields which create a Boson (a Fermion) at position 
$x$ in the Lowest Landau Level: 
\begin{equation}
\begin{array}{rcl}
\Phi _{b}^{+}\left( \vec{x}\right) &=&\sum_{s}\left\langle \vec{x}%
|s\right\rangle a_{s}^{+} \\
& & \\
\Phi _{f}^{+}\left( \vec{x}\right) &=&\sum_{s}\left\langle \vec{x}%
|s\right\rangle b_{s}^{+}
\\ & &
\end{array} \label{5}
\end{equation}
We define the
field $A^{+}\left( \vec{x}\right) $ which creates the exciton (destroys a
fermion and creates a boson) at position $\vec{x}:$%
\begin{equation}
A^{+}\left( \vec{x}\right) =1/\sqrt{N}\Phi _{b}^{+}\left( \vec{x}\right)
\Phi _{f}\left( \vec{x}\right)
\end{equation}
The Fourier modes of $A^{+}\left( x\right) $ are given by:
\begin{equation}
A_{\vec{p}}^{+}=e^{ip_{x}p_{y}/2}1/\sqrt{N}%
\sum_{s}e^{-ip_{x}s}a_{s}^{+}b_{s-p_{y}}  
\label{8}
\end{equation}
In a similar way we also define the densities of Bosons and fermion
operators: 
\begin{equation}
\begin{array}{rcl}
\rho ^{b}\left( \vec{x}\right) &=&\Phi _{b}^{+}\left( \vec{x}\right) \Phi
_{b}\left( \vec{x}\right) \\
 & & \\
\rho ^{f}\left( \vec{x}\right) &=&\Phi _{f}^{+}\left( \vec{x}\right) \Phi
_{f}\left( \vec{x}\right) 
\end{array}\label{eq:9}
\end{equation}
and their Fourier transforms as in (8).\ One ends up with similar
expressions: 
\begin{equation}
\begin{array}{rcl}
\rho _{\vec{p}}^{b} &=&e^{i\vec{p}_{x}\vec{p}_{y}/2}%
\sum_{s}e^{-ip_{x}s}a_{s}^{+}a_{s-p_{y}} \\
& & \\
\rho _{\vec{p}}^{f} &=&e^{i\vec{p}_{x}\vec{p}_{y}/2}%
\sum_{s}e^{-ip_{x}s}b_{s}^{+}b_{s-p_{y}}
\\ & &
\end{array}\label{eq:10}
\end{equation}

The commutations between these fields lead to a generalization of the
magnetic translation algebra \cite{girvin}.\ The relations between the $\rho 
$ themselves are given by: 
\begin{equation}
\begin{array}{rcl}
\left[ \rho _{\vec{p}}^{b},\rho _{\vec{q}}^{b}\right] &=&\left( e^{-i\vec{p}%
\times \vec{q}/2}-e^{i\vec{p}\times \vec{q}/2}\right) \rho _{\vec{p}+\vec{q}%
}^{b} \\
& & \\
\left[ \rho _{\vec{p}}^{f},\rho _{\vec{q}}^{f}\right] &=&-\left( e^{i\vec{p}%
\times \vec{q}/2}-e^{i\vec{p}\times \vec{q}/2}\right) \rho _{\vec{p}+\vec{q}%
}^{f} \\
& & \\
\left[ \rho _{\vec{p}}^{b},\rho _{\vec{q}}^{f}\right] &=&0
\\ & &
\end{array}\label{eq:6}
\end{equation}
where $\vec{p}\times \vec{q}=p_{x}q_{y}-p_{y}q_{x}$.\ The relation between
the $\rho $ and the $A^{+}$ are: 
\begin{equation}
\begin{array}{rcl}
\left[ \rho _{\vec{p}}^{b},A_{\vec{q}}^{+}\right] &=&e^{-i\vec{p}\times \vec{%
q}/2}A_{\vec{p}+\vec{q}}^{+} \\
& & \\
\left[ \rho _{\vec{p}}^{f},A_{\vec{q}}^{+}\right] &=&e^{-i\vec{p}\times \vec{%
q}/2}A_{\vec{p}+\vec{q}}^{+}  
\\ & &
\end{array}\label{eq:7}
\end{equation}

When we express the commutator between $A_{\vec{p}}$ and $A_{\vec{q}}^{+}$
in terms of $\rho _{h}$ the normal ordered form of $\rho _{f}$  
the commutator takes form: 
\begin{equation}
\left\{ A_{\vec{p}},A_{\vec{q}}^{+}\right\} =\delta _{\vec{p},\vec{q}%
}+1/N\left( e^{-i\vec{p}\times \vec{q}/2}\rho _{\vec{p}-\vec{q}}^{b}-e^{i%
\vec{p}\times \vec{q}/2}\rho _{\vec{p}-\vec{q}}^{h}\right) 
\label{eq:tintin}
\end{equation}
Assuming that the coefficient of $1/N$ is an operator of order one, up to a $%
1/N$ correction this commutator is equal to the usual commutator between
creation and annihilation operators.\ This will be our main approximation in
the following.

There is a natural representation of the operators $\rho _{\vec{p}}^{b}\rho
_{\vec{p}}^{h}$ and $A_{\vec{p}}^{+}$ in terms of creation and annihilation
operators obeying Fermionic commutation relations: $\left\{ c_{\vec{p}},c_{%
\vec{q}}^{+}\right\} =\delta _{\vec{p},\vec{q}}$.\ It is given by: 
\begin{equation}
\begin{array}{rcl}
\rho _{\vec{p}}^{b} &=&\sum_{r}e^{-i\vec{p}\times \vec{r}/2}c_{%
\vec{p}+\vec{r}}^{+}c_{\vec{r}} \\
& & \\
\rho _{\vec{p}}^{h} &=&\sum_{r}e^{i\vec{p}\times \vec{r}/2}c_{%
\vec{p}+\vec{r}}^{+}c_{\vec{r}} \\
& & \\
A_{\vec{p}}^{+} &=&c_{\vec{p}}^{+}  
\\ & &
\end{array}\label{eq:17}
\end{equation}
It is easy to verify that the relations (\ref{eq:6}, \ref{eq:7}) are satisfied
by this representation.
We have made the assumption that the $1/N$ correction can
be neglected in the commutator (\ref{eq:tintin}) and replace the field
$A_{\vec{p}}^{+}$ by $c_{\vec{p}}^{+}$ in the following.

The Hamiltonian which governs the dynamics of the model is given by the
projection of the interaction potential energy on the lowest Landau
level.\
\begin{equation}
H=1/2\Omega \sum_{\vec{q}}\tilde{V}\left( \vec{q}\right)
\rho ^{t}\left( \vec{q}\right) \rho ^{t}\left( -\vec{q}\right)
\end{equation}
where $\tilde{V}\left( \vec{q}\right) =e^{-q^{2}/2}\int d^{2}x
V\left( \vec{x} \right) e^{i\vec{q}\vec{x}}.$

If we use the Fermionic representation of $\rho ^{t},$ it can be
expressed in a more conventional form of an interacting Fermion
Hamiltonian. 
The ground state
energy $E_{0}$ can be evaluated in a Hartree--Fock
approximation.\ Denoting $n\left( p\right) $ the ground state distribution $%
\left( n\left( p\right) =1\;\mathrm{if}\;p<k_{f},n\left( p\right) =0\mathrm{%
\ if}\right. $
 
\noindent%
%
$\left. p>k_{f}\right) $ one obtains:
\begin{equation}
E_{0}=1/2\Omega \sum_{\vec{p},\vec{q}}\tilde{V}\left( \vec{q}\right) 4\sin
^{2}\left( \vec{p}\times \vec{q}/2\right) n\left( \vec{p}\right) \left(
1-n\left( \vec{p}-\vec{q}\right) \right)
\end{equation}
Using this expression we can determinate the appropriate Landau parameters
in this approximation \cite{pines}:
\begin{equation}
\begin{array}{rcl}
\varepsilon \left( p\right) &=&1/2\Omega \sum_{\vec{q}}\tilde{V}\left( \vec{q%
}\right) 4\sin ^{2}\left( \vec{p}\times \vec{q}/2\right) \left( 1-n\left(
\vec{p}-\vec{q}\right) -n\left( \vec{p}+\vec{q}\right) \right) \\
& & \\
f\left( \vec{p},\vec{q}\right) &=&-1/\Omega \tilde{V}\left( \vec{p}-\vec{q}%
\right) 4\sin ^{2}\left( \vec{p}\times \vec{q}/2\right)
\\ & &
\end{array}\label{eq:12}
\end{equation}
 
From this dispersion relation we can deduce the fermi velocity at the Fermi
momentum $k_{f}=\sqrt{2},v_{f}=7.5\;10^{-2}$.\ If we compare the effective
mass $m^{\ast }=k_{f}/v_{f}$ with the ``bare mass'' $m_{0}=2\pi $ defined by
the curvature of the dispersion relation at zero momentum $\varepsilon
\left( p\right) =p^{2}/2m_{0}$ one has approximately $m^{\ast }/m_{0}\approx
3$.\ Note that due to the lack of Galilean invariance of the theory, there
is no relation between the mass and the Landau parameter $F_{1}.$

The homogeneous transport equation which follows from the Landau theory is
given by \cite{pines}:
\begin{equation}
\left( -s+\cos \left( \theta \right) \right) \hat{n}\left( \theta \right)
+1/2\pi \cos \left( \theta \right) \int_{0}^{2\pi }d\theta F\left( \theta
-\theta ^{\prime }\right) \hat{n}\left( \theta ^{\prime }\right) =0
\end{equation}
where the fluctuation of the quasiparticle distribution takes the form
\begin{equation}
\delta n\left( \vec{p},\vec{r},t\right) =\delta \left( \varepsilon \left(
\vec{p}\right) -\mu \right) \hat{n}\left( \theta \right) e^{i\vec{q}\vec{r}%
-\omega t} 
\end{equation}
$s=\omega /qv_{f}$ and $\theta $ labels a point on the Fermi surface.\ The
coefficient $F\left( \theta \right) =\Omega m^{\ast }f\left( \vec{p},\vec{p}%
^{\prime }\right) /2\pi ,$ where $\vec{p},\vec{p}^{\prime }$ are two momenta
on the Fermi surface making an angle $\theta $ with each other.

The two first Fourier modes $\left( F_{n}=\int d\theta /2\pi F\left( \theta
\right) \right) $ of $F\left( \theta \right) $ 
are both less than zero, furthermore, when $p_{f}/\sqrt{2}%
=.985\;F_{0}$ becomes less than $-1$ and the system becomes unstable (The
compressibility is negative) and $F_{1}=-.5$.\ 

Let us compare the expression of
the static form factor we obtain with some theoretical predictions.\ The
form factor $S\left( q\right) $ is defined as: 
\begin{equation}
S\left( q\right) =< \rho ^{t}\left( \vec{q}\right) \rho ^{t}\left( -\vec{q}\right)>
\end{equation}

We can obtain it by setting ${\tilde V} = 2 \delta ^{\vec q}$ in (10).
For $q$ small $S\left( q\right) $ behaves as $2k_{f}q^{3}/3\pi $ instead of $%
2q/\pi k_{f}$ in a Fermi liquid.\ The $q^{3}$ behavior of the form factor
agrees with the numerical predictions \cite{haldane}. 
As for the quantum Hall effect form factor 
\cite{girvin} there is a $q^{2}$ reduction with respect to the normal Fermi
liquid behavior at small $q.$

For $q>2k_{f},2\pi S\left( q\right) $ goes to $k_{f}^{2}$ as $k_{f}^{2}\left(
1-2j_{1}\left( qk_{f}\right) /qk_{f}\right) $ where $j_{1}$ denotes the
Bessel function (we use a normalization of $\rho _{q}$ for which there is no
exponential behavior of $S\left( q\right) $ at large $q$ unlike in \cite
{girvin}).\ The limiting value $k_{f}^{2}$ is an exact result related to the
Casimir operator of the $\rho $ algebra.\ This
prediction also indicates the limitation of the present model to the $\nu =1$
Bosonic case since in the physical $\nu =1/2$ Fermionic case the limiting
value should be $1/2$ (The general limiting value of is $1+\nu \left\langle
P_{ij}\right\rangle $ where $\left\langle P_{ij}\right\rangle $ denotes the
expectation value of the permutation operator $+1$ for Bosons, $-1$ for
Fermions \cite{haldane}).\ This $S\left( q\right) $ does not reproduce a
cusp singularity at $2k_{f}$ which is indicated by the simulations \cite
{haldane}.

We can gain a better understanding of the preceding result by comparing the
expression of the form factor with the response function evaluated in the
quasistatic limit: $\left( s=\omega /v_{f}q\ll 1\right) .$\ In this limit
one can interpret the system as a $2D$ Fermi liquid consisting of dipoles.\
At the Fermi surface the dipole vector of a quasiparticle with a momentum
equal to $k_{f}$ is given by $d_{i}=\varepsilon _{ij}k_{j}$.\ Let $\rho
_{q}=\sum_{k}c_{k+q}^{+}c_{k}$ denote the Fourier modes of the dipole
density.\ A scalar potential $\phi \left( \vec{r},t\right) $ acts on the
system through an interacting Hamiltonian: 
\begin{equation}
H_{e}=\sum_{\vec{q}}\int d\omega \rho _{-\vec{q}}\vec{q}.\vec{d}\phi \left( 
\vec{q},\omega \right) e^{-i\omega t} 
\end{equation}
where $\phi \left( \vec{q},\omega \right) $ is the Fourier transform in
space and time of $\phi \left( \vec{r},t\right) $.\ In the long wavelength
limit $\left( q\ll k_{f}\right) $ the dipole-vector can be replaced by its
value at the Fermi surface and the net effect is to replace the interaction
Hamiltonian by the usual coupling to a scalar potential: 
\begin{equation}
H_{e}=\sum_{\vec{q}}\int d\omega \rho _{-\vec{q}}^{t}\phi \left( \vec{q}%
,\omega \right) e^{-i\omega t}  \label{29}
\end{equation}
where $\rho _{\vec{k}}^{t}=\sum_{\vec{q}}\left( \vec{k}\times \vec{q}\right)
c_{\vec{k}+\vec{q}}^{+}c_{\vec{q}}$ denotes the long wavelength limit of the
total density operator.\ The response function is defined as: 
\begin{equation}
\chi \left( q,\omega \right) =\left\langle \rho ^{t}\left( \vec{q},\omega
\right) \right\rangle /\phi \left( \vec{q},\omega \right) .
\end{equation}
To evaluate it we use the transport equation in the presence of the external
force due to the scalar potential.\ It reads: 
\begin{equation}
\begin{array}{rcl}
\left( -s+\cos \left( \theta \right) \right) \hat{n}\left( \theta \right)
+\cos \left( \theta \right) \hat{n}\left( \theta ^{\prime }\right) d\theta
^{\prime }/2\pi\\
& & \\
\qquad \qquad \qquad \qquad \qquad \qquad =\left( qk_{f}\right) \sin \left(
\theta \right) \cos \left( \theta \right) \phi \left( \vec{q},\omega \right)
\\ & &
\end{array}\label{eq:19}
\end{equation}
We expend the solution of this equation in powers of $s$ and make use of the
fact that: $\left\langle \rho ^{t}\left( \vec{q},\omega \right)
\right\rangle =\sum_{\vec{p}}\left( \vec{p}\times \vec{q}\right) \delta
n\left( \vec{p}\right) $.\ Form the first term we deduce the static response
function: 
\begin{equation}
\chi \left( q,0\right) =-\left( qk_{f}\right) ^{2}\nu \left( 0\right)
/2\left( 1+F_{1}\right)
\end{equation}
where $\nu \left( 0\right) =m^{\ast }\Omega /2\pi $ is the density of states
on the Fermi surface.

Note the $q^{2}$ dependence which is different from the usual Fermi liquid
behavior and that $F_{1}$ appears instead of $F_{0}$.\ This result does not
satisfy the usual compressibility sum rule because the scalar potential does
not see the quasiparticles as elementary but rather as dipoles.\ As a
result, it does not deform the Fermi sea symmetrically.

The next order gives an imaginary contribution which is related to the
dynamical Form factor \cite{pines}.
One deduces the following expressions: 
\begin{equation}
S\left( \vec{q},\omega \right) =\left( 2m^{\ast }\omega /\left( 2\pi \right)
^{2}v_{f}q\right) \left( k_{f}q/1+F_{1}\right) ^{2}
\end{equation}
The first factor is the free fermion result.\ The factor $\left(
k_{f}q\right) ^{2}$ was predicted at the Hartree--Fock approximation and
originates from the fact that dipoles couple much more weakly to the scalar
potential as ordinary quasiparticles.\ Finally the many-body effects
renormalize the Hartree--Fock contribution by $\left( 1+F_{1}\right)
^{-2}\approx 4$ (instead of $\left( 1+F_{0}\right) ^{-2})$ in the usual
Fermi liquid case.

Let $J$ denote the quasiparticle current at the Fermi surface.\ In the
dipole theory we define a modified current $J^{t}$ in such a way that the
variation of the kinetic energy with the time is proportional to $\vec{J}%
^{t}\left( \vec{x}\right) =\vec{d}.\vec{\nabla}J\left( \vec{x}\right) $
where $\vec{d}$ denotes the dipole vector.\ The computation of the
conductivity then proceeds as for the response function and the net effect
of this redefinition is to renormalize the Fermi liquid result by the factor 
$\left( k_{f}q\right) ^{2}$.
The physical interpretation is simple: The only
quasiparticles which can radiate must travel with the group velocity of the
electric field.\ When the group velocity is small compared to $v_{f}$ they
lie at the two points of the Fermi surface with a Fermi momentum
perpendicular to $\vec{q}$.\ Since the electric field is not constant in the
direction of $\vec{q},$ it couples to the dipole vector of the
quasiparticles and the system radiates when $\vec{E}$ is parallel to their
velocity.
Unfortunately, this argument
does not produce the longitudinal conductivity
($\vec E$ and $\vec q$ in the $x$ direction) which is observed in the experiments \cite{willet}%
 but a transverse conductivity $\sigma _{\perp }\left( \vec{q}\right) 
=e^{2}\sqrt{2}q/\pi $.


\newpage

\centerline{\epsffile{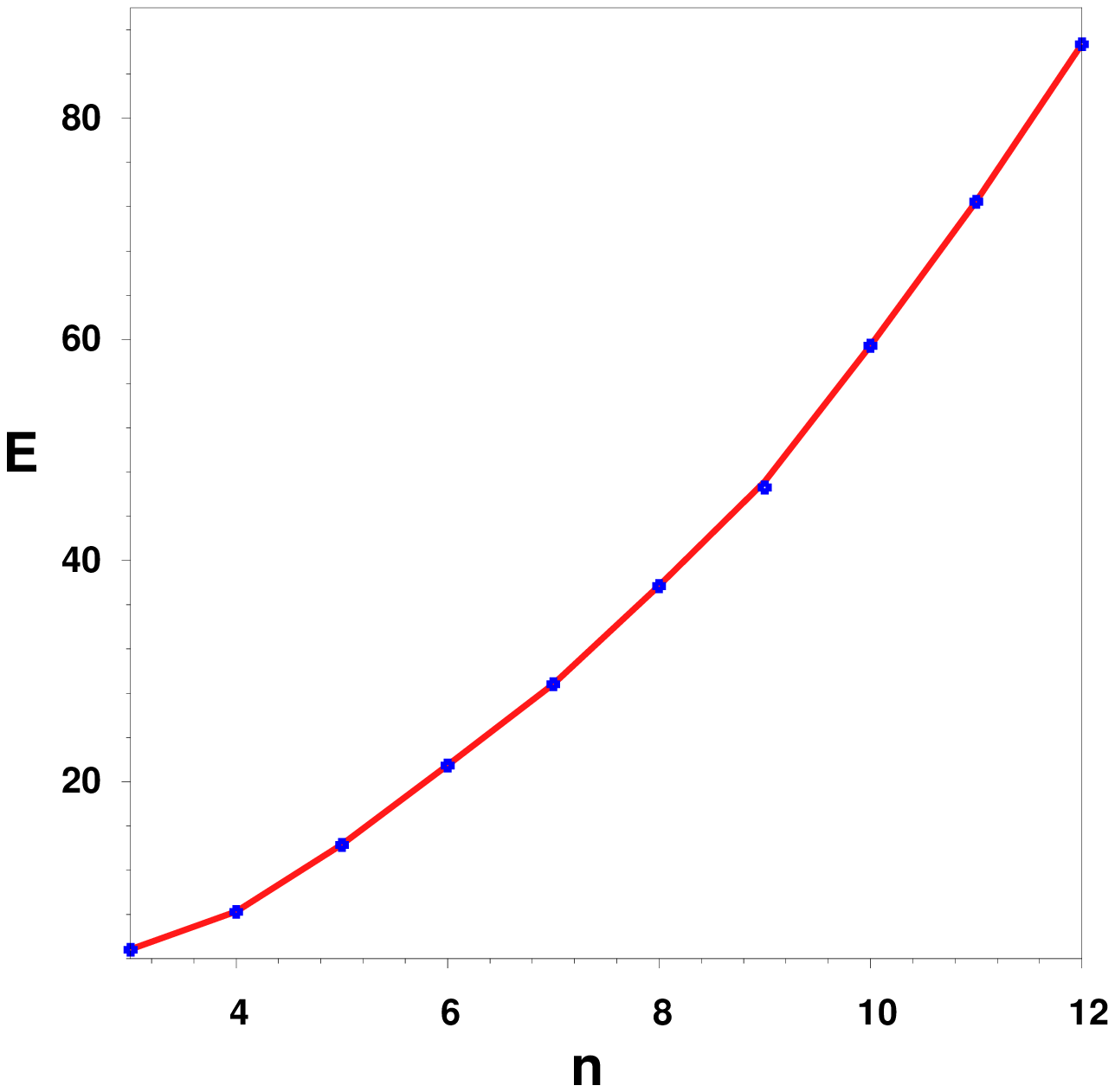}}

{{\it Fig.1.}~We plot the ground state energy of Bosons on a sphere at
filling factor $\nu =1.$\ The Bosons interact through a $\delta $
repulsive potential.\

We plot the energy as a function of the number of bosons $n\left( 1\leq
n\leq 12\right) .$\ As a function of $n,$ the energy is roughly linear
by pieces with a slope breaking at each perfect square
$\left( n=4,9\right) .$\ This indicates that the bosons have a similar
ground state energy as a system of $n$ free fermions on a sphere
\textit{without} magnetic field.}

\newpage
\centerline{\epsffile{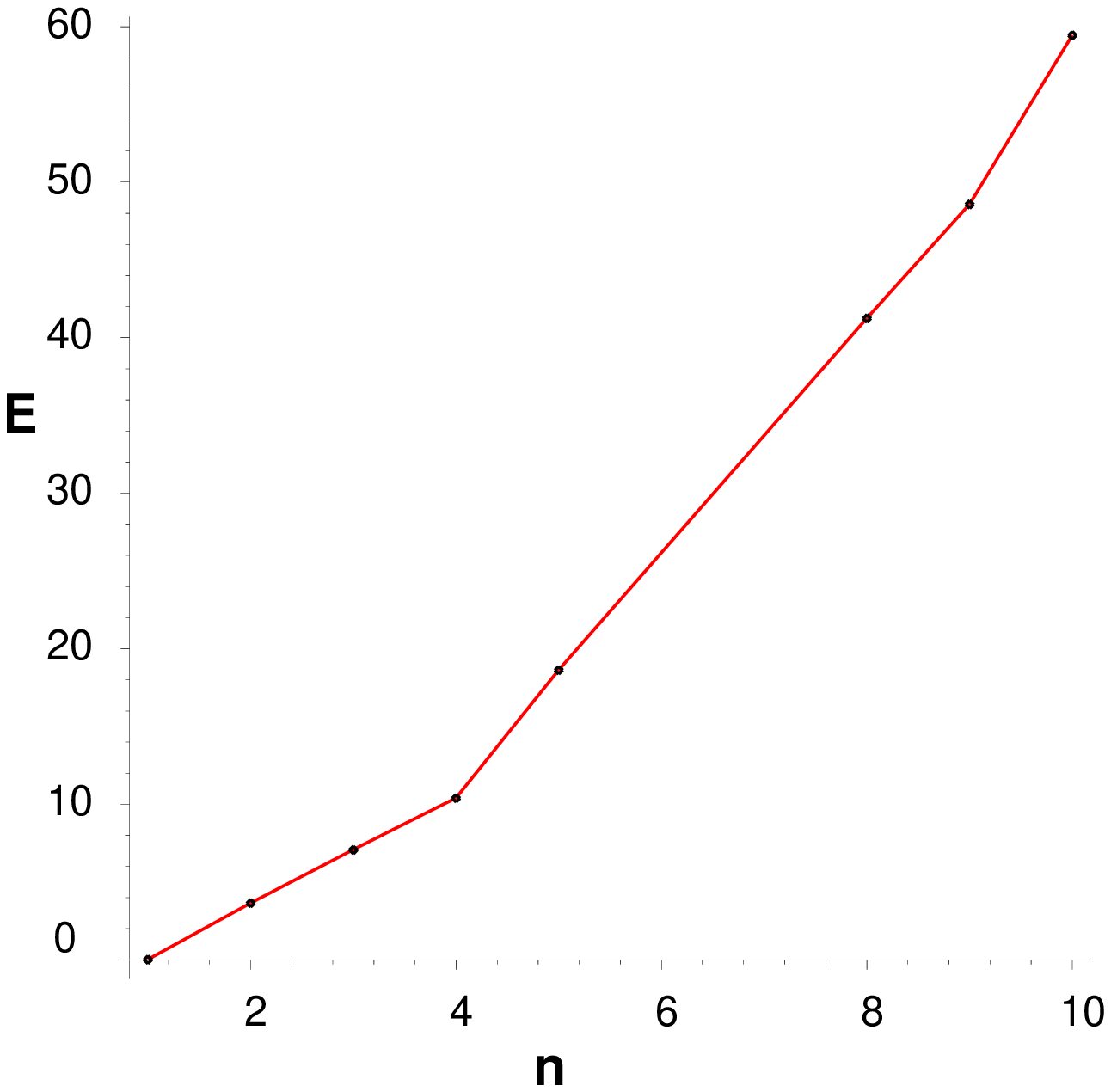}}

{{\it Fig.2.}~We consider a system of 10 particles interacting on a sphere with a
magnetic field at $\nu =1.$\ the interaction potential is a $\delta$
function as in fig.1.

These particles are splitted into $n$ bosons and $10-n$ fermions and we plot
the ground state energy as a function of $n.$\ We see that even more
convincingly as in figure 1 the energy behaves a if there were $n$ free
fermions without magnetic field.\ The interpretation is that the boson binds
to the fermionic hole to form a quasi-free bound state.\ The striking fact
is that this feature remains true even where there is no fermion left $%
!\left( n=10\right) .$
}

\end{document}